\documentclass[letterpaper,twocolumn,american,showpacs,prl,aps,superscriptaddress]{revtex4}
\usepackage[latin1]{inputenc}
\usepackage{bm}
\usepackage{multirow}
\usepackage{amssymb}
\usepackage{amsbsy}
\usepackage{amsmath}
\usepackage{stmaryrd}
\usepackage{graphicx}
\usepackage{subfigure}
\usepackage{epsfig}
\usepackage{hyperref}
\makeatletter
\usepackage{pifont}
\makeatother

\begin{document}

\title{Dynamical typicality of quantum expectation values}

\author{Christian Bartsch}
\email{cbartsch@uos.de}
\affiliation{Fachbereich Physik, Universit\"at Osnabr\"uck,
             Barbarastrasse 7, D-49069 Osnabr\"uck, Germany}

\author{Jochen Gemmer}
\email{jgemmer@uos.de}
\affiliation{Fachbereich Physik, Universit\"at Osnabr\"uck,
             Barbarastrasse 7, D-49069 Osnabr\"uck, Germany}

\date{\today}

\begin{abstract}
We show that the vast majority of all pure states featuring a common expectation value
of  some generic observable at a given time will yield very similar expectation values of the same observable at any later time. This is meant to apply to Schr\"odinger type dynamics in high dimensional Hilbert spaces. As a consequence individual dynamics of expectation values are then typically well described by the ensemble average.
Our approach is based on the Hilbert space average method.
We support the analytical investigations with numerics obtained by exact diagonalization
of the full time-dependent Schr\"odinger equation for some pertinent, abstract Hamiltonian model. Furthermore, we discuss the
implications on the applicability of projection operator methods with respect to initial states, as well as on irreversibility in general. 
\end{abstract}


\maketitle

In its broadest sense the term typicality may be explained as follows \cite{lebowitz1993, lebowitz2007}: If a set of  states specified by some common feature (e.g., drawn according to the same distribution, sharing the same energy, etc.) yields a very
narrow distribution of some other feature (e.g. some observable, etc.), then there is typicality. The concept of typicality as key to the occurrence of standard statistical equilibrium behavior (as opposed to ergodicity, mixing, etc.) especially in quantum 
mechanics has recently been established in various works \cite{goldstein2006,popescu2006,reimann2007,reimann2008}.

One important implementation has been presented in \cite{goldstein2006,popescu2006}. There it is shown that a majority of pure states of a compound system from some narrow energy shell yield almost the same 
reduced density matrix for a small subsystem. This happens to be the
canonical equilibrium state in case of weak interactions and generic
environment spectra \cite{goldstein2006}.

In another typicality-based investigation \cite{reimann2007} it is demonstrated that states drawn 
according to a certain type of probability distribution in Hilbert space feature
very similar  quantum expectation values (QEV's) of generic observables. Even more detailed results exist for the special case of a uniform distribution of normalized, pure states, i.e., a distribution which is invariant under all unitary transformations in Hilbert space \cite{comment1}. Using the Hilbert space average method (HAM) \cite{gemmer2003,gemmer2004} one finds that the ``Hilbert space
average'' ($\mathrm{HA}$), i.e., the average  of the QEV's  of an observable $D$ w.r.t. the above distribution, is given by 
\begin{equation}
\mathrm{HA}\lbrack\langle\psi\vert D\vert\psi\rangle\rbrack = \frac{\mathrm{Tr}\{ D\} }{n}=c_1
\label{HA}
\end{equation}
and the corresponding ``Hilbert space variance'' ($\mathrm{HV}$) by 
\begin{eqnarray}
\mathrm{HV}\lbrack\langle\psi\vert D\vert\psi\rangle\rbrack &:=&\mathrm{HA}\lbrack \left( \langle\psi\vert D\vert\psi\rangle -\mathrm{HA}\lbrack\langle\psi\vert D\vert\psi\rangle\rbrack \right)^2\rbrack \nonumber\\
&&\!\!\!\!\!\!\!\!\!\!\!\!\!\!\!\!\!\!\!\!\!\!\!\!\!\!\!\!\!\!\!\!\!\!\!\!\!\!\!\!\!\!\!\!\!\!\!\!\!\!\!\!\!\!=\frac{1}{n+1}\left( \frac{\mathrm{Tr}\{ D^2 \}}{n} - \left( \frac{\mathrm{Tr}\{ D\}}{n}\right)^2 \right)
=\frac{1}{n+1}\left( c_2 - c_1^2 \right)\!\! ,
\label{HV}
\end{eqnarray}
cf. \cite{gemmer2004}. Here $n$ denotes the dimension of the corresponding Hilbert space and $c_i:=\mathrm{Tr}\{ D^{i}\}/n$ describes the $i$-th moment of the spectrum of $D$. Thus $c_2-c_1^2$ is the spectral variance of $D$. Throughout this paper we focus on observables the low spectral moments of which do not change (significantly) under physically reasonable ``upscaling''. Pertinent examples are, e.g., a component of a specific spin in a system which is upscaled by adding more and more (interacting) spins, the occupation number of some momentum mode in an interacting many-particle system which is upscaled to comprise more and more momentum modes, or in general any local variable embedded in a growing system. For any such observable one may conclude from (\ref{HV}) that the Hilbert space variance, i.e., the width of the above distribution of QEV's vanishes with growing dimension $n$. In this sense bound observables in large systems yield typical QEV's.

In this paper we turn towards the typicality of dynamics of QEV's.
In short, we demonstrate in the paper at hand that pure states from a set $\{ \vert\phi\rangle\}$ featuring 
a common QEV of some observable $A$ at some time $t$, i.e.
$\langle\phi\vert A(t)\vert\phi\rangle = a$, most likely yield very similar QEV's at any later time, i.e. 
$\langle\phi\vert A(t+\tau)\vert\phi\rangle \approx \langle\phi'\vert A(t+\tau)\vert\phi'\rangle $ (with $ \vert\phi\rangle,\vert\phi\rangle'$ both being states from the above set).
We present some analytical derivations based on the HAM, in particular on Eqs.~(\ref{HA} and \ref{HV})
and we additionally support the results with numerical calculations.
Finally, we discuss what consequences arise for the validity of  projection operator methods 
(Nakajima-Zwanzig (NZ), etc. \cite{breuer2006,breuer2007,fick1990}) w.r.t. initial states and the corresponding inhomogeneities. Furthermore, we comment on the irreversibility of QEV's corresponding to individual pure states.

We specify our considered observable $A$ only by the above mentioned moments, $c_i$, and specialize without substantial loss of generality to observables which are trace-free, $c_1=0$, and normalized to  $c_2=1$. Furthermore we require the  $c_i$ with $i=2,...,8$ to be of the order $1$.  
Next, we introduce an ensemble of pure states $\vert\phi\rangle$ which is 
characterized as follows: All its states must feature 
the same QEV of the observable $A$, $\langle\phi\vert A\vert\phi\rangle = a$, must be  normalized ($\langle\phi\vert\phi\rangle=1$), 
and uniformly distributed otherwise.
That means
the ensemble has to stay invariant under all unitary transformations
in Hilbert space that leave the expectation value of $A$ unchanged, 
i.e. those transformations that commute with $A$, or, concretely, transformations of the 
form $e^{iB}$, with $\lbrack B,A\rbrack =0$.
This specifies the most general ensemble consistent with the restriction that all its states should yield a given
$a$.

For the following calculations we further introduce some kind of ``substitute'' ensemble $\{\vert\omega\rangle\}$, which 
is much easier to handle.
As will be shown below, this ensemble approximates
the exact ensemble $\{\vert\phi\rangle\}$ described above very well for large  Hilbert spaces.

The ensemble $\{\vert\omega\rangle\}$ is generated by
\begin{equation}
\vert\omega\rangle=(1/\sqrt{1+d^{2}})(1+dA)\vert \psi \rangle \ ,
\label{substens}
\end{equation}
where $\vert\psi\rangle$ are pure states drawn from a uniform distribution of normalized states without further restriction as described above (\ref{HA}). $d$ is some small parameter which describes 
the deviation from the ``equilibrium'' ensemble $\{ \vert\psi\rangle\}$. Since it is essentially the operator $A$ itself that generates $\{\vert\omega\rangle\}$ from the entirely uniform distribution, $\{\vert\omega\rangle\}$ is invariant under the above uniform transformations that leave $a$ invariant.

The construction (\ref{substens}) allows for an evaluation of moments of the distribution of $\langle\omega\vert C\vert\omega\rangle$ based on results on moments of the distribution of $\langle\psi\vert D\vert\psi\rangle$, or concretely
\begin{eqnarray}
&&\mathrm{HA}\lbrack \langle\omega\vert C\vert\omega\rangle^i\rbrack =\mathrm{HA}\lbrack \langle\psi\vert D\vert\psi\rangle^i \rbrack \nonumber\\
&& \mbox{with}\quad D=\frac{1}{1+d^2}(1+dA)C(1+dA)\ .
\label{omegaexpec}
\end{eqnarray}
(Of course the average on the l.h.s. corresponds to the substitute ensemble $\{\vert\omega\rangle\}$ while the average on the r.h.s is based on the completely uniform ensemble $\{\vert\psi\rangle\}$). Exploiting this, average and variance of  $\langle\omega\vert C\vert\omega\rangle$ may be evaluated with the help of (\ref{HA},\ref{HV}).

To assure that the ensemble $\{\vert\omega\rangle\}$ indeed approximates the ensemble $\{\vert\phi\rangle\}$,
in the limit of large $n$, we evaluate the following four 
quantities
\begin{eqnarray}
\mathrm{HA}\lbrack\langle\omega\vert\omega\rangle\rbrack, \quad\quad && \quad\quad \mathrm{HA}\lbrack\langle\omega\vert A\vert\omega\rangle\rbrack \ , \nonumber\\
\mathrm{HV}\lbrack\langle\omega\vert\omega\rangle\rbrack, \quad\quad && \quad\quad \mathrm{HV}\lbrack\langle\omega\vert A(t)\vert\omega\rangle\rbrack \ ,
\label{quantdef}
\end{eqnarray}
where $A(t)$ denotes the time dependence according to the Heisenberg picture.
(For clarity: the results are given in  Eqs.~(\ref{normsave}), (\ref{normsvar}), (\ref{expecave}) and (\ref{expecbound}).)

The states $\vert\omega\rangle$ are not exactly normalized
which would render them unphysical, of course. However, one finds from (\ref{HA}) and (\ref{omegaexpec}) 
(by implementing $C=1$) that
\begin{equation}
\mathrm{HA}\lbrack \langle\omega\vert\omega\rangle \rbrack =1\ .
\label{normsave}
\end{equation}

By exploiting (\ref{HV}) and (\ref{omegaexpec}) one finds analogously for the variance
\begin{equation}
\mathrm{HV}\lbrack \langle\omega\vert\omega\rangle \rbrack = \frac{1}{n+1}\cdot\frac{4d^2+4d^3c_3+d^4(c_4-1)}{(1+d^2)^2} \ .
\label{normsvar}
\end{equation}
As defined above, the $c_i$ are of the order $1$, i.e. the $\mathrm{HV}$ of the norms scales with $1/n$ and
becomes small for large Hilbert spaces. Therefore, the vast majority of the states $\vert\omega\rangle$ are approximately normalized for 
large $n$.

The average of the QEV's of  $A$  w.r.t. the ensemble $\{\vert\omega\rangle\}$  (which is meant to correspond to the above $a$) is calculated by
exploiting (\ref{HA}) and (\ref{omegaexpec}) (by implementing $C=A$)
\begin{equation}
\mathrm{HA}\lbrack\langle\omega\vert A\vert\omega\rangle\rbrack=\frac{2d+d^{2}c_3}{1+d^{2}} \ .
\label{expecave}
\end{equation}
That is, the mean QEV can be adjusted through the choice of the parameter $d$. However,
the replacement ensemble is restricted on expectation values not too far away from zero  (i.e. the average expectation value of the ``equilibrium'' ensemble $\{\vert\psi\rangle\}$) because by sweeping through all possible $d$ not all possible expectation
values up to the maximum eigenvalue of $A$ are reachable.

The evaluation of $\mathrm{HV}\lbrack\langle\omega\vert A(t)\vert\omega\rangle\rbrack$ turns out to be somewhat more complicated,
since we, in general, cannot fully diagonalize the Hamiltonian and thus do not know $A(t)$ in detail. However, we are 
able to perform an estimation for an upper bound.  
For this purpose we make use of the Hilbert Schmidt scalar 
product for complex matrices defined as
$(X,Y):=\mathrm{Tr}\{X^{\dagger} Y\}$.
Thus, one can formulate a Cauchy-Schwarz inequality of the form
\begin{equation}
\mathrm{Tr}\{X^{\dagger} Y\} \leq \sqrt{\mathrm{Tr}\{X^{\dagger}X\}\mathrm{Tr}\{Y^{\dagger}Y\}}\ .
\label{CS}
\end{equation}
Particularly, one obtains $\mathrm{Tr}\{ A(t)A \} \leq \mathrm{Tr}\{ A^{2} \}$.
Evaluating  $\mathrm{HV}\lbrack\langle\omega\vert A(t)\vert\omega\rangle\rbrack$ based on (\ref{HV}) and (\ref{omegaexpec}) (by implementing $C=A(t)$), realizing that $\mathrm{Tr}\{D\}^2$ is always positive and repeatedly applying
(\ref{CS}) yields the inequality  
\begin{eqnarray}
&&\mathrm{HV}\lbrack\langle\omega\vert A(t)\vert\omega\rangle\rbrack\leq \frac{1}{n+1}\cdot \nonumber\\
&&\!\!\!\!\!\!\frac{1+4d\sqrt{c_4}+6d^{2}c_4+4d^{3}\sqrt{c_4}\sqrt[4]{c_4 c_8}+d^{4}\sqrt{c_4 c_8}}{(1+d^{2})^{2}}\ .
\label{expecbound}
\end{eqnarray}
Again, since the $c_i$ are of the order $1$, the upper bound deceases with $1/n$. Thus, the variance
(\ref{expecbound}) becomes small for large Hilbert spaces, just like the variance of the norms (\ref{normsvar}).
This result yields two major direct implications. 

First, if one evaluates
(\ref{expecbound}) at $t=0$, one finds that the majority of the states $\vert\omega\rangle$ feature approximately the same QEV of the observable $A$ for large $n$. From this property together with the result that the states $\vert\omega\rangle$ are nearly normalized one concludes that the replacement ensemble $\{\vert\omega\rangle\}$ indeed approximates the exact ensemble $\{\vert\phi\rangle\}$ very well for large Hilbert spaces (with $a=\mathrm{HA}\lbrack\langle\omega\vert A\vert\omega\rangle\rbrack$ as given in (\ref{expecave})).

Second, the upper bound from (\ref{expecbound}) is valid for any time $t$.
Thus, for large enough systems, the dynamical curves for $a_\omega (t):=\langle\omega\vert A(t)\vert\omega\rangle$ of the vast majority
of pure states from the initial ensemble $\{\vert\omega\rangle\}$ are very close to each other and thus to the evolving ensemble average at any time $t$. Due to the similarity of $\{\vert\omega\rangle\}$ and $\{\vert\phi\rangle\}$ this should also hold  true for the ``exact'' ensemble $\{\vert\phi\rangle\}$. 
Thus, there is a typical evolution for the expectation values $\langle\phi\vert A(t)\vert\phi\rangle$ or, to rephrase, there is ``dynamical typicality''.
This statement represents the main result of this paper.
Particularly, this typicality is independent of the concrete form of the dynamics,
which may be a standard exponential decay into equilibrium or something completely different.

\begin{figure}[htb]
\centering
\hspace{-1.0cm}
\subfigure[]{\label{exp}
\includegraphics[width=6.5cm]{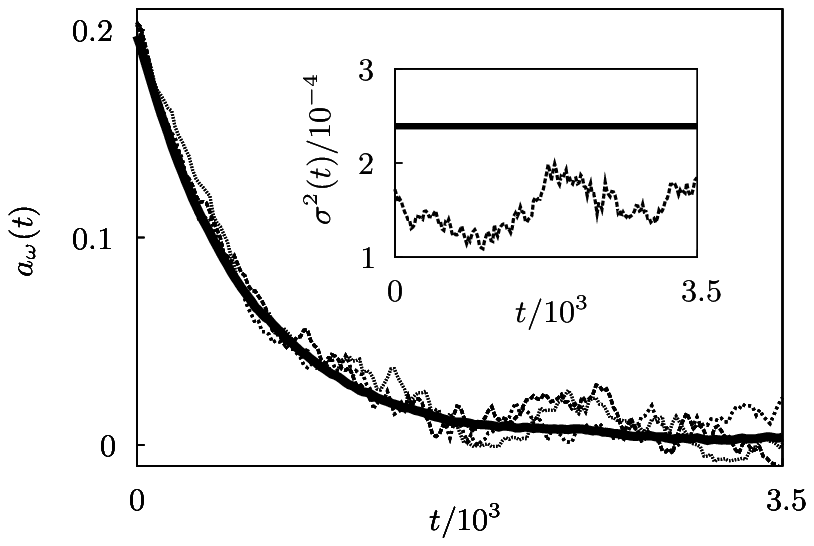}}
\hspace*{-1.0cm}
\subfigure[]{\label{gauss}
\includegraphics[width=6.5cm]{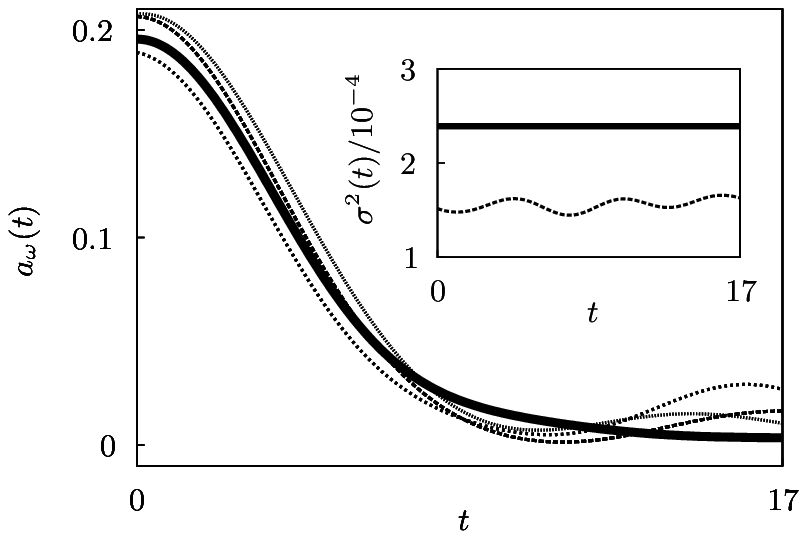}}
\subfigure[]{\label{const}
\includegraphics[width=6.5cm]{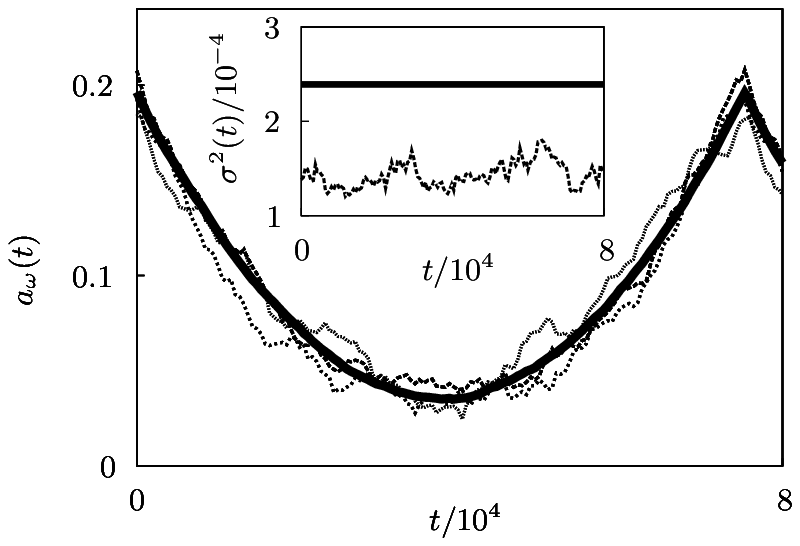}}
\hspace{1.0cm}
\caption{Dynamics of some expectation values corresponding to a set of initial states $\{\vert\omega\rangle\}$, i.e., $\langle \omega\vert A(t)\vert\omega\rangle=a_{\omega}(t)$. The set is characterized by an ``common'' initial expectation value, i.e., all $\vert\omega\rangle$ yield $\langle \omega\vert A(0)\vert\omega\rangle \approx 0.2$. The figures illustrate that the individual evolutions typically stay close to the average over the set (solid line). The insets show evolutions of the variances $\sigma^2(t)$ which stay accordingly small and remain below some analytically predicted upper bound (solid line), cf. (\ref{expecbound}). The subfigures correspond to different Hamiltonians, generating different archetypical types of dynamics: a.) exponential relaxation, b.) non-exponential relaxation, c.) non-relaxing.} 
\label{graphs}
\end{figure}

In the following we visualize these predictions for a model
quantum system described by a Hamiltonian of the form $H=H_0+V$, 
where $H_0$ is some unperturbed Hamiltonian with equidistant energy eigenvalues (number of states: $n=6000$, level spacing: $\Delta E=8.33\cdot 10^{-5}$, $\hbar$ set to $1$),
and $V$ some possibly but not necessarily small interaction.
$A$ is chosen as diagonal in the eigenbasis of $H_0$ with equally many, randomly placed elements $1$ and $-1$. This is in accordance with the already mentioned
conditions on the moments $c_i$ ($c_1=0$ , $c_2=1$).
However, this specific form of the Hamiltonian and the observable $A$ is not crucial for
the main results of this paper concerning the typicality of expectation values. 
It is just an example for the numerical illustrations. 
We further calculate the dynamical curves of $a_{\omega(t)}$ as generated by three concrete interactions $V$
that are chosen to represent different generic types of dynamical behavior.
Thus, the matrix elements of $V$ in the eigenbasis of $H_0$ (and $A$) are taken as 
(i.) random complex Gaussian numbers with $\overline{V_{ij}}=0, \overline{|V_{ij}|^2}= 2.25\cdot 10^{-8} $ (small perturbation),
(ii.) random complex Gaussian numbers with $\overline{V_{ij}}=0, \overline{|V_{ij}|^2}= 6.25\cdot 10^{-6} $ (strong perturbation),
(iii.) all identical constants, i.e., $V_{ij}^2=2.25\cdot 10^{-8}$.
We obtain the dynamics of the $a_{\omega} (t)$ by numerically solving the full time-dependent Schr\"odinger equation).
These interactions give rise to three archetypical evolutions
(i.) an exponential decay into equilibrium (Fig.~\ref{exp}),
(ii.) non-exponential decay into equilibrium (Fig.~\ref{gauss}),
(iii.) no decay to equilibrium at all, even in the limit of many states and weak perturbations (Fig.~\ref{const}).
The precise reasons for the emergence of these dynamics are beyond the scope of this text (for more details see \cite{bartsch2008}). For a clarification of the term "relaxation" in this context refer to \cite{comment2}. 

In each of the figures the dynamics of $a_{\omega}(t)$ for three different pure states of the ensemble $\{\vert\omega\rangle\}$ as initial states are displayed. Furthermore, the evolution of the ensemble average is shown (averaged over $100$ states). To numerically generate $\{\vert\omega\rangle\}$, the  real and imaginary parts of the amplitudes of the states $\vert\psi\rangle$ (cf. (\ref{substens})) w.r.t. the  eigenbasis of $H_0$ are drawn as independent 
Gaussian numbers, similar to the Gaussian Adjusted Projected ensemble (GAP) (see \cite{goldsteinstat2006,reimannstat2008}). $d$ is chosen as $0.1$.
Essentially, one finds that, for all three interactions, the dynamical curves of the pure states are close to the average curve for all
displayed times $t$.
Furthermore, the insets in Figs.~\ref{exp},\ref{gauss},\ref{const} show the numerical variance $\sigma^2 (t)$ of the $a_{\omega}(t)$ (calculated for $100$ states), which turns out to be always
smaller than the upper bound given in (\ref{expecbound}). Thus, all these numerics illustrate and back up our previously derived analytical results.

We now address further implications.
The mean QEV, i.e., essentially $a$, can alternatively be reformulated using the notion of a density matrix
as usually done in the framework of projection operator formalisms
\begin{equation}
a=\mathrm{HA}\lbrack\langle\omega\vert A\vert\omega\rangle\rbrack=\mathrm{HA}\lbrack\mathrm{Tr}\{ A\vert\omega\rangle\langle\omega\vert\}\rbrack=\mathrm{Tr}\{A\ \mathrm{HA}\lbrack\vert\omega\rangle\langle\omega\vert\rbrack\}\ .
\end{equation}
The $\mathrm{HA}\lbrack\vert\omega\rangle\langle\omega\vert\rbrack$ takes the role of the density matrix. Further evaluation gives
(using the ``substitute'' ensemble $\{\vert\omega\rangle\}$) (see \cite{goldstein2006,gemmer2004})
\begin{equation}
\mathrm{HA}\lbrack\vert\omega\rangle\langle\omega\vert\rbrack = \frac{1+2dA+d^{2}A^{2}}{n(1+d^2)}\ .
\end{equation}
For ensembles close to equilibrium, i.e., small $d$, which is fulfilled in the examples presented here, 
one can neglect the terms which grow quadratically in $d$. 
In this case, the density matrix takes approximately the same form as the initial state which is
often used  in projection operator calculations which aim at determining the dynamics of
expectation values like $a(t)$ (\cite{bartsch2008}). There, for reasons given below, the (mixed) initial state is simply taken to be $\rho(0)=1/n+cA$ such that $c=a(0)$.
That means, correct dynamical results from the projection operator methods based on the above initial state describe the dynamics of the
ensemble average of $\{ \vert\omega\rangle \}$.

From this point of view some consequences on the applicability of 
projection operator theories (NZ, time-convolutionless, Mori formalism etc.), which are standard tools
for the description of reduced dynamics, arise. These methods
have in common the occurrence of an inhomogeneity in the central equations of motion that typically has to be neglected in order to solve them. Generally, the inhomogeneity depends on the true initial state, it, however, vanishes if the true initial state indeed is of some specific form determined by the pertinent projector \cite{breuer2007, weiss1999, fick1990}. For the above mentioned case the above $\rho(0)$ is exactly of that form, which means the dynamics of the ensemble are equal to the dynamics generated by the pertinent projected equation of motion without the inhomogeneity. However, the evolution of the ensemble is typical, this implies that the inhomogeneity, as generated by most of the true initial states, should be negligible. 

On the other hand, there are investigations in the field of open quantum systems, e.g., \cite{pechukas1994} and \cite{romero2004}, suggesting that the true initial states may have an utterly crucial influence on the dynamics, such that, e.g., some correlated initial states may yield  projected dynamics which are entirely different from the ones obtained by corresponding product states.  

Nevertheless, to rephrase, the results of this paper indicate that in the limit of large (high dimensional) systems the inhomogeneity should 
become more and more irrelevant in the sense that the statistical weight of initial states, which yield an inhomogeneity that substantially changes the solution of the projected equation of motion, should decrease to zero. Note that this does not contradict the concrete results of \cite{pechukas1994} and \cite{romero2004}.

The above results also shed some light on the relation of the apparently irreversible dynamics of QEV's to the, in some sense, reversible dynamics of the underlying Schr\"odinger equation. If a mean QEV as generated by some initial non-equilibrium ensemble (pertinent density matrix) relaxes to equilibrium \cite{comment2} (which can often be reliably shown \cite{breuer2007}) , then for the majority of the individual states that form the ensemble, the corresponding individual QEV's will relax to equilibrium in the same way. Thus, for the relaxation of the QEV's, the question whether or not the initial ensemble truly exists is largely irrelevant. Of course, there may be individual initial states giving rise to QEV evolutions that do not (directly) relax to equilibrium, but, to repeat, for high dimensional systems, their statistical weight is low.

Support by the Deutsche Forschungsgemeinschaft through the Graduiertenkolleg 695 is gratefully acknowledged.


\begin{thebibliography}{19}
\bibitem[{com({\natexlab{a}})}]{comment1}
\bibinfo{note}{One may consider all below investigations as applying to a part
  of the Hilbert space that corresponds to a certain energy region of the
  system rather than to the full Hilbert space. That amounts to microcanonical
  scenarios of a given finite temperatur rather than infinite temperature.}
\end{thebibliography}
\end{document}